\documentclass[letterpaper]{article} % DO NOT CHANGE THIS
\usepackage{aaai2026}  % DO NOT CHANGE THIS
\usepackage{times}  % DO NOT CHANGE THIS
\usepackage{helvet}  % DO NOT CHANGE THIS
\usepackage{courier}  % DO NOT CHANGE THIS
\usepackage[hyphens]{url}  % DO NOT CHANGE THIS
\usepackage{graphicx} % DO NOT CHANGE THIS
\usepackage{natbib}  % DO NOT CHANGE THIS AND DO NOT ADD ANY OPTIONS TO IT

\usepackage{algorithm}
\usepackage{algorithmic}
\usepackage{times}
\usepackage{helvet}
\usepackage{courier}
\usepackage{graphicx}
\usepackage{caption}
\usepackage{subcaption}

\usepackage{booktabs}
\usepackage{makecell}
\usepackage{multirow}
\usepackage{amsthm}
\usepackage{etoolbox}
\usepackage{algorithm}
\usepackage{algorithmic}
\usepackage{placeins}
\usepackage{amsmath}
\usepackage{amssymb}
\usepackage[table,xcdraw]{xcolor}

\definecolor{lightgray}{rgb}
{0.95,0.95,0.95}
\usepackage{helvet}
\usepackage{courier}
\usepackage{newfloat}
\usepackage{listings}
\usepackage{newfloat}
\usepackage{listings}
\DeclareCaptionStyle{ruled}{labelfont=normalfont,labelsep=colon,strut=off} % DO NOT CHANGE THIS
\floatstyle{ruled}
\newfloat{listing}{tb}{lst}{}
\floatname{listing}{Listing}
\title{Self-Supervised One-Step Diffusion Refinement for Snapshot Compressive Imaging}
\author{
    Shaoguang Huang\textsuperscript{\rm 1}, 
    Yunzhen Wang\textsuperscript{\rm 1}, 
    Haijin Zeng\textsuperscript{\rm 2}\textsuperscript{$\ast$}, 
    Hongyu Chen\textsuperscript{\rm 1}, 
    Hongyan Zhang\textsuperscript{\rm 1}\thanks{Corresponding author}
}

\affiliations{
    \textsuperscript{\rm 1}School of Computer Science, China University of Geosciences, Wuhan, China
    \\
    \textsuperscript{\rm 2}Department of Psychiatry, Harvard University, Cambridge, USA   \\ \{huangshaoguang, chenhongyu, zhanghongyan\}@cug.edu.cn, yunzhen1110@163.com, haijin.zeng2018@gmail.com

}

\usepackage{bibentry}
\begin{document}

\maketitle

\begin{abstract}
Snapshot compressive imaging (SCI) captures multispectral images (MSIs) using a single coded two-dimensional (2-D) measurement, but reconstructing high-fidelity MSIs from these compressed inputs remains a fundamentally ill-posed challenge. While diffusion-based reconstruction methods have recently raised the bar for quality, they face critical limitations: a lack of large-scale MSI training data, adverse domain shifts from RGB-pretrained models, and inference inefficiencies due to multi-step sampling. These drawbacks restrict their practicality in real-world applications.
In contrast to existing methods—which either follow costly iterative refinement or adapt subspace-based embeddings for diffusion models (e.g., DiffSCI, PSR‑SCI)—we introduce a fundamentally different paradigm: a self-supervised One-Step Diffusion (OSD) framework designed specifically for SCI. The key novelty lies in using a single-step diffusion refiner to correct an initial reconstruction, eliminating iterative denoising entirely while preserving generative quality.
Moreover, we adopt a self-supervised equivariant learning strategy to train both the predictor and refiner directly from raw 2-D measurements, enabling generalization to unseen domains without ground-truth MSI. To further address limited MSI data, we design a band-selection–driven distillation strategy that transfers core generative priors from large-scale RGB datasets, effectively bridging the domain gap.
Extensive experiments confirm that our approach sets a new standard—yielding PSNR gains of 3.44dB, 1.61dB, and 0.28dB on the Harvard, NTIRE, and ICVL datasets respectively—while cutting reconstruction time by 97.5\%, from 8.9s to just 0.22s per image. This leap in efficiency and adaptability makes our method a major advancement in SCI reconstruction—both accurate and practical for real-world deployment. The codes and extended version of this paper are available at  \url{https://github.com/shhuang-1767/DiFASCI}.
\end{abstract}

\section{Introduction}

Multispectral images (MSIs) capture rich spectral information within more spectral bands than conventional RGB images, enabling to distinguish between different materials that might appear identical in the RGB image. Thus, MSIs find applications in environmental monitoring~\cite{2, 10769516}, land cover classification~\cite{3}, anomaly detection ~\cite{4} and material identification~\cite{1}. Traditional multispectral imaging systems acquire MSIs by spatial or temporal scanning, leading to a long acquisition time. Driven by compressive sensing theory, snapshot compressive imaging (SCI) systems~\cite{11} capture 2D measurements in a single shot, enabling efficient and cost-effective MSI acquisition with the aid of MSI reconstruction algorithms.

\begin{figure}[t]
    \centering
    \includegraphics[width=1.0\linewidth]{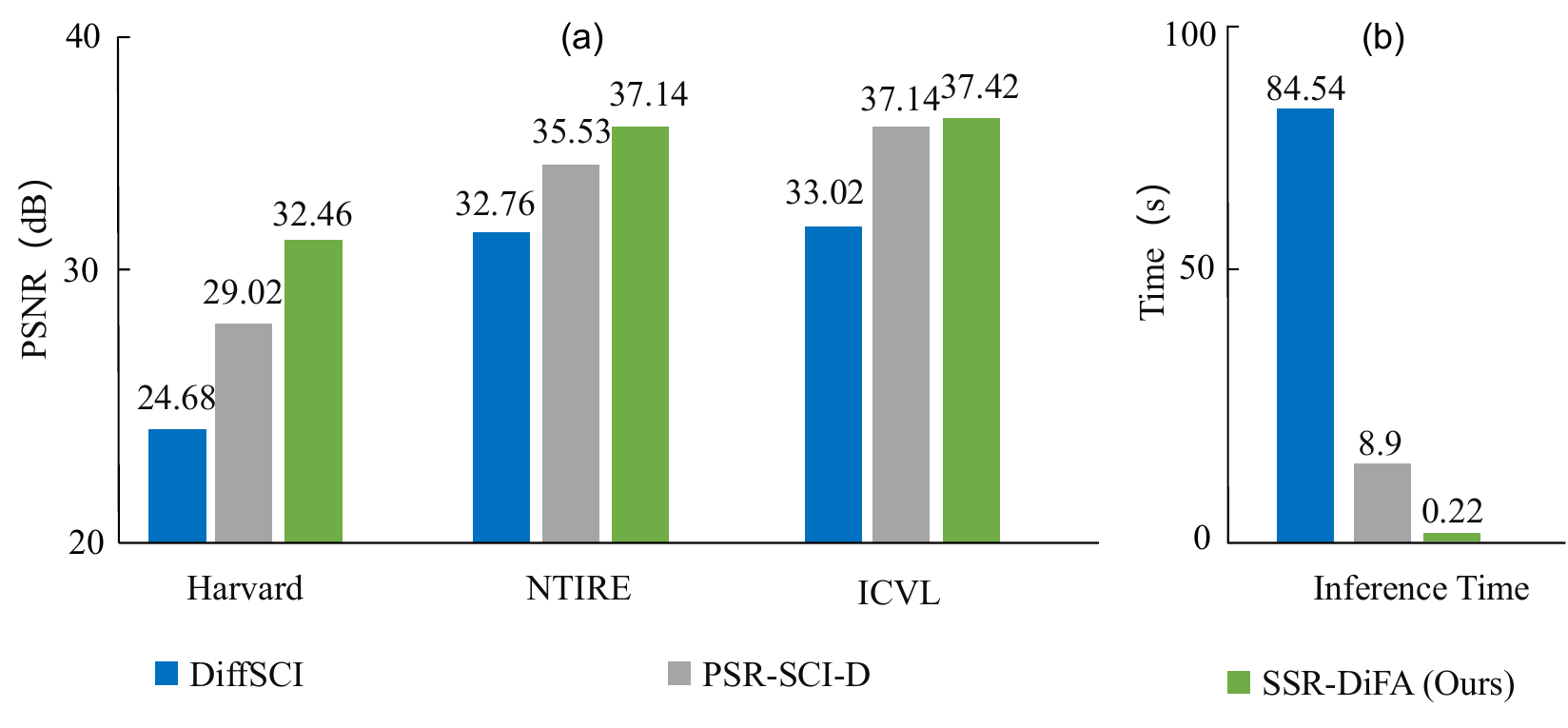}
    \vspace{-8mm}
    \caption{Comparison with the latest diffusion-based SCI reconstruction methods on zero-shot datasets Harvard, NTIRE and ICVL in terms of (a) PSNR and (b) inference time.}
    \label{fig:teaser}
\end{figure}

Restoring a 3-D MSI from the 2-D measurement in SCI systems is an ill-posed inverse problem. Existing SCI restoration methods can be categorized into model-based and deep-learning-based approaches. The traditional model-based methods typically solve the inverse problem within a convex optimization framework with hand-crafted image priors, including sparsity, low-rank and local smoothness. They enjoy good theory guarantees but require manual parameter tuning. Current deep-learning-based approaches include plug-and-play (PnP) based~\cite{45, zeng2020cviu}, end-to-end (E2E) based ~\cite{zhang2024improving} and deep unfolding-based methods~\cite{zhang2024dual}. PnP methods integrate pre-trained denoising networks into the traditional optimization-based framework. Due to the fixed pre-trained denoising model, the performance of PnP methods is often limited. E2E methods learn a mapping function from 2-D measurement to 3-D MSI through supervised learning by using paired data. However, they often neglect the physical characteristics of SCI systems and thus are less interpretable and robust to different hardware systems. Deep unfolding-based methods unfold a convex optimization process by using multistage neural networks to restore MSI, resulting in an interpretable neural network architecture. However, in the inadequate sampling areas of 2-D measurement, existing methods often struggle to obtain accurate restoration with fine details \cite{zeng2025spectral}.

Diffusion model has yielded remarkable success in generating content from RGB images \cite{ho2020denoising}. Recent studies \cite{wu2025latent, pan2024diffsci, zeng2025spectral} leverage its generative capacity to enhance the quality of MSI in SCI, achieving promising results~\cite{zeng2024unmixing}. DiffSCI~\cite{pan2024diffsci} employs a pre-trained diffusion model on large RGB datasets as a generative denoiser within the PnP framework. Zeng et al. propose a diffusion-based two-stage framework where the high-frequency component of MSIs is refined by a fine-tuned diffusion model in the low dimensional subspace~\cite{zeng2025spectral}. Another recent work~\cite{wu2025latent} adopts a latent diffusion model to generate image priors, which are fed into a deep unfolding network for high-quality MSI reconstruction. Despite the encouraging results, the following issues remain: 1) The generation ability of the used diffusion model is limited due to the domain shift between RGB training data and MSI. Although the methods in~\cite{zeng2025spectral,wu2025latent} fine-tune or retrain the diffusion models with paired measurements and MSIs, the scarcity of paired MSI data (e.g., merely a few thousand MSIs versus hundreds of millions of RGB images, and sometimes completely inaccessible) leads to suboptimal performance in the derived models' generative capabilities. 2) The iterative nature of diffusion models necessitates significant sampling time, a limitation that becomes more pronounced in MSIs due to the high dimensionality of MSI.

% Thus, the key challenges we would like to address are: 1) how to train a customized diffusion model from scratch for MSI reconstruction task without using paired MSIs. 

To address these problems, we propose an efficient self-supervised one-step diffusion refinement framework to enhance the details of MSI. We leverage an existing SCI reconstruction network to generate an initial prediction of MSI from the input 2-D measurement and use a one-step diffusion model to generate an MSI residual for refinement. To ensure robust performance on unseen datasets, the SCI reconstruction network is co-trained with the one-step diffusion model. The coarse-to-fine strategy enables the use of a compact diffusion network by focusing on residual generation, which is significantly simpler than reconstructing the full MSI. To achieve fast one-step generation without compromising quality compared to multi-step diffusion models, we propose a spectral compression distillation loss for knowledge transfer from RGB images to MSI. Additionally, to address the lack of paired MSIs, we introduce a self-supervised learning paradigm based on the equivariant imaging framework, enabling network training solely with 2D measurements. The main contributions can be summarized as follows:
\vspace{-5pt}
\begin{itemize}
% DiFASCI employs a coarse-to-fine strategy in a unified framework, first generating a deterministic prediction by using any existing SCI reconstruction methods and then refining it through an efficient one-step diffusion model. Experimental results demonstrate that our approach outperforms recent state-of-the-art methods in both visual quality and reconstruction accuracy.
% DiFASCI generates high-frequency details for MSIs， simplifying the modeling process while ensuring fast inference speeds (approximately 0.2s for a 256×256×28 image). This is the first application of one-step diffusion in SCI.
\item [1)]
We propose an efficient self-supervised one-step diffusion refinement framework for SCI reconstruction. Our approach requires no paired MSIs, making it highly suitable for real-world applications where only 2D measurements are accessible. The framework also exhibits excellent generalization capabilities, as it can be flexibly combined with any E2E and deep unfolding-based SCI reconstruction networks.

\item [2)]
We design a one-step diffusion adaptation (DiFA) module to enhance the details of the reconstructed MSIs from an initial predictor. It simplifies the whole modeling process while facilitating faster inference speed compared with the existing diffusion-based SCI reconstruction methods.

\item [3)]
To leverage the prior knowledge from pre-trained multi-step diffusion models, we propose spectral compression distillation in RGB space, enabling fast one-step MSI generation without compromising quality.

% We propose a spectral compression distillation loss with the guidance of the pre-trained multi-stage diffusion module on large-scale RGB datasets, which ensures fast one-step generation for MSI without sacrificing generation quality.
%

\item [4)]
Due to the lack of paired MSIs, recovering MSI is impossible solely from the 2-D measurement with a data fidelity loss. We develop an additional self-supervised training loss based on the transformation invariant assumption on MSI, allowing the model to be trained only with the compressed measurements.

% \item [4)] We evaluate our method on simulated and real datasets. Our method outperforms the latest diffusion SCI reconstruction methods as shown in Fig. \ref{fig:teaser} and other SOTAs as shown in Table. \ref{tab:average}.
\end{itemize}

\section{Related Works}
\subsection{MSI Restoration Methods}
Existing SCI reconstruction frameworks mainly include model-based and deep learning-based methods. Model-based methods~\cite{20,zeng2020hyperspectral} restore MSI within an optimization-based framework. To cope with the ill-posed inverse problem, they typically impose hand-crafted constraints such as sparsity, low-rank, graph Laplacian and total variation on the restored data by taking into account image priors. These methods are interpretable but often show low reconstruction quality and speed. 

End-to-end and deep unfolding are two representative deep learning-based frameworks for SCI reconstruction. The end-to-end approaches~\cite{fu2021coded,xiong2017hscnn,meng2020end, miao2019net, cheng2022recurrent, cai2022mask, cai2022coarse, zhang2024improving} directly learn a mapping function on paired MSIs to restore MSIs from the 2-D compressed measurements. Deep unfolding methods~\cite{zhang2021learning, zhang2022herosnet, huang2021deep, meng2020gap, ma2019deep, wang2020dnu, 
cai2022degradation,
li2023pixel,
dong2023residual,
zhang2024dual} unfold the iterative optimization-based reconstruction process into a multi-stage network, where each stage corresponds to one iteration of the optimization algorithm. However, these methods typically involve minimizing $\ell_1$ or $\ell_2$ based pixel loss and thus encounter the regression-to-mean issue, facing limitations in reconstructing high-frequency details \cite{wu2025latent}.

Deep generative models such as the DDPM~\cite{ho2020denoising, nichol2021improved} directly sample from the posterior, where a set of specific candidates can be generated from the learned posterior instead of calculating the average of all possible solutions, showing great potential for MSI reconstruction. Recently, attempts \cite{wu2025latent, pan2024diffsci, zeng2025spectral} have been made to leverage the generative ability of diffusion models to enhance the reconstruction of MSI. Specifically, DiffSCI~\cite{pan2024diffsci} employs a pre-trained RGB image denoising diffusion model as the denoiser in a PnP framework. Zeng et al. develop a two-stage SCI reconstruction method PSR-SCI~\cite{zeng2025spectral} where a diffusion model is used to enhance the high-frequency component of the MSI in a lower subspace derived with unmixing. Another recent work~\cite{wu2025latent} integrates a latent diffusion model to generate clean image priors for a deep unfolding network, resulting in high-quality data reconstruction.

\subsection{ResShift}
ResShift~\cite{yue2024efficient} is a diffusion model starting with a prior distribution centered on the low-quality (LQ) image, which allows the high-quality (HQ) image to be iteratively recovered from its LQ counterpart instead of Gaussian white noise. The forward process can be expressed as follows:
\begin{equation}
    q(x_t\mid x_0,y)=\mathcal{N}\left(x_t;x_0+\eta_t (y - x),\kappa^2\eta_t\mathbf{I}\right),
    \label{eq:forward}
\end{equation}
where $y$ is a LR image, $x_{0}$ is a HR image,  $\kappa$ is a hyper-parameter for the noise variance and $\eta_t$ is a shifting sequence, which is monotonically increasing and satisfied $\eta_{1} \to 0$ and $\eta_{T} \to 1$. Based on this forward process, the reverse process will commence from the initial state with rich information in LQ images to perform denoising. The formula is as follows:
\begin{equation}
  \begin{aligned}
  p_\theta(x_{t-1}|x_t,y)&=\mathcal{N}\left(x_{t-1}\bigg|\frac{\eta_{t-1}}{\eta_t}x_t\right.+ \frac{\alpha_t}{\eta_t}x_{0},\kappa^2\frac{\eta_{t-1}}{\eta_t}\alpha_t\mathbf{I}\bigg),
  \label{eq:reverse}
  \end{aligned}
\end{equation}
where $\alpha_t = \eta_t - \eta_{t-1}$ for $t > 1$ and $\alpha_1 = \eta_1$, and an initial state is $x_T = y+\kappa\sqrt{\eta_T}\epsilon$, where $\epsilon \sim \mathcal{N}(\mathbf{0},\mathbf{I})$. In the reverse process in Eq. (\ref{eq:reverse}), $x_{0}$ is predicted by a trainable deep neural network $f_{\theta}$. A non-Markovian reverse process exists which produces deterministic sampling and is reformulated as:
\begin{equation}
    q(x_{t-1}\mid x_t,x_0,y)=\delta(k_tx_0+m_tx_t+j_ty),
    \label{eq:deterministic}
\end{equation}
where $\delta$ is the unit impulse and $k_{t}$, $x_{t}$, $j_{t}$ and $m_t$ are derived from $\eta_{t}$. More details can be found in~\cite{wang2024sinsr}. Eq. (\ref{eq:deterministic}) indicates a deterministic mapping from $x_{T}$ to $x_{0}$, expressed as $F_{\theta}(x_{T}, y)$.

Especially, the reverse process in one-step generation can be formulated as:
\begin{equation}
    x_{0} = f_{\theta}(x_{T}, y, T).
    \label{one-step}
\end{equation}
where the HQ image $x_{0}$ is generated by the trainable network $f_{\theta}$ directly, and $T$ is the total timestep of ResShift.
\section{The Proposed Method}
\subsection{Problem Definition}
\subsubsection{3.1.1 Degradation Model of CASSI.}
One of the representative SCI systems is the Coded Aperture Snapshot Spectral Compressive Imaging (CASSI) system~\cite{15}. It captures 3-D spectral data in a single snapshot by encoding spatial and spectral information through a coded aperture and reconstructing it from compressed 2-D measurements.

Let $\mathcal{X}\in\mathbb{R}^{H\times W\times B}$ denote a 3-D MSI and $\mathcal{Y}\in\mathbb{R}^{H\times W+d\times (B-1)}$ be the 2-D measurements, where $H$, $W$, $d$, and $B$ are the MSI’s height, width, shifting step, and total number of wavelengths, respectively. To obtain a concise equation for the imaging process, let $\mathbf{y} \in R^{n}$ with $n = H (W + d(B - 1))$ represent the vectorized measurement, $\mathbf{x} \in R^{nB}$ be the vectorized shifted MSI and $\mathcal{H}\in\mathbb{R}^{n\times nB}$ be the mask. Then, we obtain the imaging process in CASSI ~\cite{meng2020end,ma2019deep}:
\begin{equation}\label{eq:CASSI}
   \mathbf{y} = \mathcal{H} \mathbf{x} + \mathbf{n},
\end{equation}
where $\mathbf{n} \in R^{n}$ represents the noise on measurement. 

Then, the next step is to obtain $\mathbf{x}$ given the obtained measurement $\mathbf{y}$ and $\mathcal{H}$.

\begin{figure*}[!h]
\begin{center}
\includegraphics[width=\textwidth]{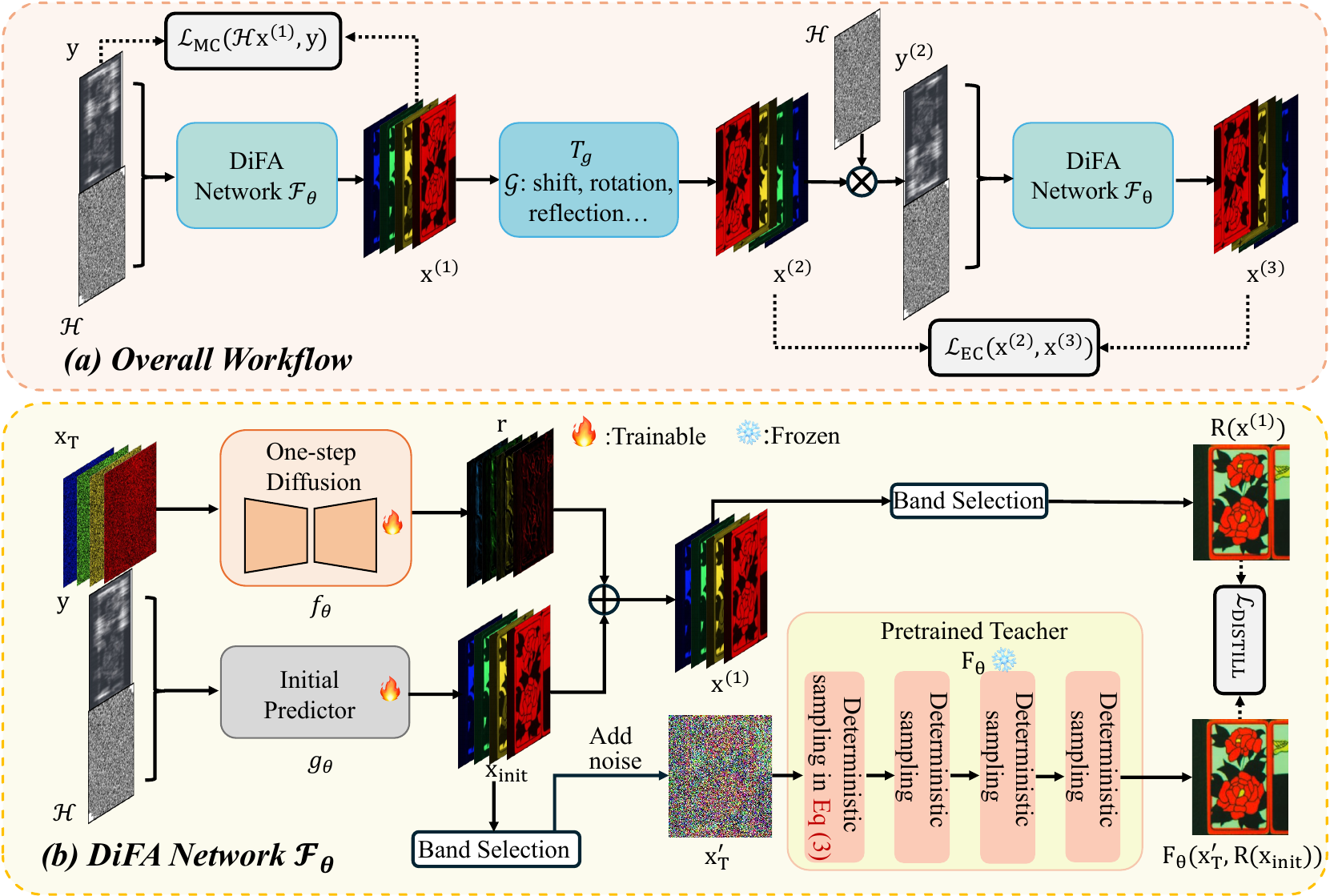}
\end{center}
    \vspace{-4.5mm}
    \caption{(a) Workflow for our self-supervised training strategy. The measurement $\mathbf{y}$ and mask $\mathcal{H}$ are initially input into DiFA Network $\mathcal{F_{\theta}}$, resulting in the recovered MSI $\mathbf{x^{(1)}}$. Next, a series of transformations $T_{g}$ containing shift, rotation, reflection, etc. are applied to $\mathbf{x^{(1)}}$ to produce $\mathbf{x^{(2)}}$. The MSI $\mathbf{x^{(2)}}$ is then modulated again by the mask $\mathcal{H}$ to obtain the compressed measurement $\mathbf{y ^{(2)}}$, which is finally input into $\mathcal{F_{\theta}}$ to obtain the re-reconstructed MSI $\mathbf{x^{(3)}}$. (b) DiFA Network, on the one hand, the measurement $\mathbf{y}$ and mask $\mathcal{H}$ are initially input into pre-trained reconstruction network $g_{\theta}$ to get the initial predictor $\mathbf{x}_{\mathrm{init}}$, the residual of MSI $\mathbf{r}$ is generated from noise $\mathbf{x}_{T}$ by one-step diffusion $f_{\theta}$. Then, the refined image $\mathbf{x^{(1)}}$ is obtained by adding the residual $\mathbf{r}$ to the initial prediction $\mathbf{x}_{\mathrm{init}}$. On the other hand, both $\mathbf{x}_{\mathrm{init}}$ and $\mathbf{x^{(1)}}$ are converted to RGB images through band selection $R(\cdot)$. A distillation loss is developed to leverage the prior knowledge of pre-trained multistep diffusion models.
    }
    \label{fig:framework}
\end{figure*}

\subsubsection{3.1.2 Challenges with Diffusion Model for SCI.}
Leveraging the generative power of diffusion models offers a promising solution to improve the detail of MSIs in SCI reconstruction. However, 1) directly integrating the existing pre-trained diffusion models, which are typically trained with RGB datasets, into the SCI reconstruction task is infeasible due to the band mismatch. 2) Training a diffusion model from scratch for MSI is challenging due to the limited MSI. The total amount of paired MSI in benchmark datasets is around only a thousand, which is far smaller than the RGB data. The difficulty intensifies when paired MSIs are entirely absent. 3) The inference process is time-consuming due to the high dimensionality of MSI and the substantial number of iterative sampling steps required.
\subsection{The Self-supervised One-step Diffusion Refinement for SCI}
Here, we develop a self-supervised one-step diffusion refinement framework as shown in Fig. \ref{fig:framework} (a) where a core module DiFA is designed to obtain MSI with the input of $\mathbf{y}$ and $\mathcal{H}$. As shown in Fig. \ref{fig:framework} (b), we feed $\mathbf{y}$ and $\mathcal{H}$ into an initial predictor $g_{\theta}$ to obtain an initial prediction of MSI $\mathbf{x}_{\mathbf{init}}=g_{\theta}(\mathbf{y},\mathcal{H})$. Then, we refine $\mathbf{x}_{\mathbf{init}}$ by using a residual $\mathbf{r}=f_{\theta}(\mathbf{x}_{\mathrm{init}})$ generated by a one-step diffusion model $f_{\theta}$, leading to an enhanced MSI $\mathbf{x}^{(1)}=\mathbf{x}_{\mathrm{init}}+\mathbf{r}$. The one-step diffusion model is trained from scratch by using only the measurement with self-supervised learning. Then, we compress the obtained MSI $\mathbf{x}^{(1)}$ into the RGB space and employ a distillation technique to leverage the prior knowledge of pre-trained multi-step diffusion model to guide the learning of the one-step diffusion model and the initial predictor. In response to the unavailability of paired MSIs in real applications, we develop a self-supervised loss that allows efficient model training with only 2-D measurement.
\vspace{-2pt}
\subsubsection{3.2.1 Efficient One-step Diffusion Adaption.}
Using a multi-step diffusion model for SCI reconstruction is time-consuming due to the high dimensionality of MSI and the large number of sampling steps. In addition, modeling the mapping process from 2-D measurement to 3-D MSI directly is challenging. Thus, we propose an efficient one-step diffusion adaption module DiFA, which first generates an initial prediction of MSI by an existing SCI reconstruction network and then refines it with details provided by a one-step diffusion network. Unlike the iterative noise prediction in the DDPM~\cite{ho2020denoising} reverse process, our approach employs a single-step diffusion to directly generate a clean image from an initial state, leading to a fast inference speed. 

Let $\mathcal{F}_{\theta} = g_{\theta}(\cdot) + f_{\theta}(\cdot)$ denote the DiFA as shown in Fig. \ref{fig:framework} (b), where $g_{\theta}(\cdot)$ and $f_{\theta}(\cdot)$ are the initial predictor and one-step diffusion model with $\theta$ represents the weights of neural networks. Note that our focus is not on the specific design of the initial predictor. Considering the availability of existing pre-trained SCI reconstruction neural networks, we can use an existing one as the initial predictor, which will be trained with self-supervised learning, for an initial guess of the restored MSI: $\mathbf{x}_{\mathrm{init}}=g_{\theta}(\mathbf{y})$.

Under the predict-and-refine pipeline, our OSD aims to estimate a 3-D residual $\mathbf{r}$ through a rough 3-D MSI, a similar task in the ResShift~\cite{yue2024efficient}, which recovers a HR RGB image from a LQ RGB image. However, directly employing the ResShift as our OSD is infeasible because of the disparity of bands between RGB and MSI. Retraining one from scratch for MSI is challenging due to the limited paired MSIs. Here, we design a self-supervised training method by incorporating distillation and equivariant consistency to jointly learn the OSD and initial predictor with solely the 2-D measurement. Details of this strategy are described in the following sections.

\begin{figure}[t]
    \centering    \includegraphics[width=1\linewidth]{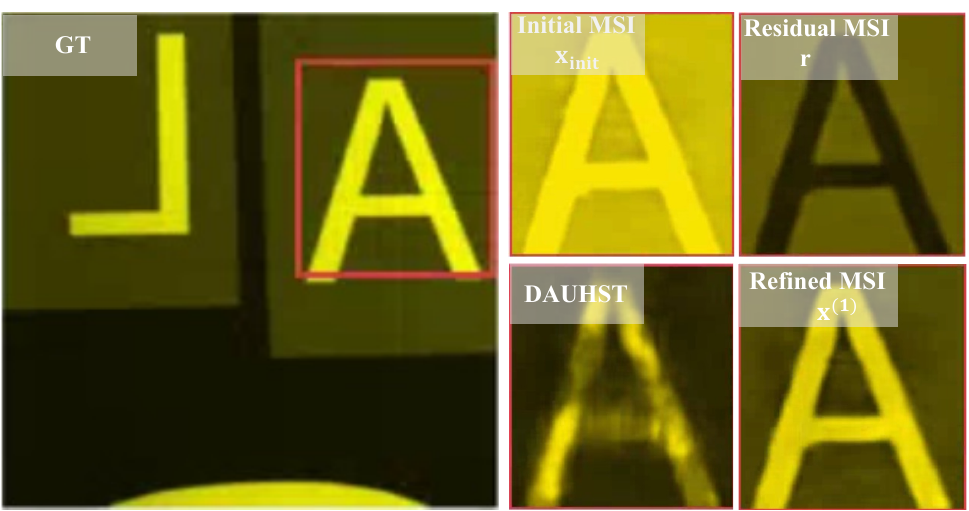}
    \vspace{-7mm}
    \caption{The refined MSI, obtained with the initial and residual MSIs, exhibits superior visual quality with finer details compared to the baseline DAUHST.}
    \label{fig:residual}
\end{figure}

Based on Eq. (\ref{one-step}), we estimate the residual $\mathbf{r}$ as follows:
\vspace{-2pt}
\begin{align}
    \mathbf{r} = f_{\theta} (\mathbf{x}_{T}, \mathbf{x}_{\mathrm{init}}, T),
\end{align}
where $\mathbf{x}_T = \mathbf{x}_{\mathrm{init}} + \kappa \sqrt{\eta_T} \epsilon$ is the initial state with $\epsilon \sim \mathcal{N}(0, 1)$ representing noise sampled from a standard normal distribution. We set $T=4$ in our paper. 
%The refined MSI, produced by the initial predictor and OSD exhibits superior visual quality with finer details and is enhanced as shown in Fig~\ref{fig:residual}.%
The visual results in Fig. \ref{fig:residual} show that the refined MSI, produced by the initial predictor and OSD, exhibits superior visual quality with finer details and enhanced sharpness.

\vspace{-3pt}
\noindent \subsubsection{3.2.2 Spectral Compression Distillation.}
The one-step diffusion model is trained from scratch for the residual estimation without using any paired MSIs. While it shows faster inference speed, its generative capability might be sacrificed. To alleviate this problem, we design a distillation method to leverage the diffusion prior from pre-trained multi-stage diffusion models on RGB data to guide the training of the one-step diffusion model and the refinement of the initial predictor. However, due to the disparity of bands between MSI and RGB images, direct distillation is infeasible. To project the MSIs into the RGB space, we select three bands at 625.0 nm, 522.5 nm, and 476.5 nm corresponding to the red, green, and blue channels, respectively. This projection process is denoted as $R(\cdot)$. Then, we perform distillation in the RGB space as follows:
% Thus, we selected the three bands at 625.0nm, 522.5nm, and 476.5nm from the MSIs to represent the red, green and blue channels, thereby projecting the MSIs into the RGB space, we denote this process as $R$. Then, we perform distillation in the RGB space as follows:
\vspace{-2pt}
\begin{equation}
    \mathcal{L}_{DISTILL}=L_{MSE}(R(\mathbf{x}^{(1)}),F_{\theta}(\mathbf{x}_{T}^{'}, R(\mathbf{x}_{\mathrm{init}}))),
\end{equation}
where $F_{\theta}$ is a pre-trained multi-step diffusion model~\cite{yue2024efficient} and $\mathbf{x}_{T}^{'}  = R(\mathbf{x}_{\mathrm{init}}) + \kappa \sqrt{\eta_T} \epsilon$ with $\epsilon \sim \mathcal{N}(0, 1)$ is the initial state in the RGB space.

\vspace{-2pt}
\subsubsection{3.2.3 Equivariant Imaging Diffusion Consistency.}
Due to the costly nature of MSI acquisition and the lack of paired MSIs for training the one-step diffusion network in real scenarios, training a DiFA network for SCI reconstruction using only compressed measurement data becomes imperative.

% \newline
\noindent \textbf{Measurement Consistency Loss.} 
Consider a naive unsupervised loss that only enforces measurement consistency: 
\begin{equation}\label{eq:MC}
    \mathcal{L}_{MC} = ||\mathbf{y} - \mathcal{H} \mathcal{F}_{\theta}(\mathbf{y}, \mathcal{H})||^{2}
                      = ||\mathbf{y} - \mathcal{H}\mathbf{x}^{(1)}||^{2},
\end{equation}
where $\mathbf{x}^{(1)}  = \mathcal{F}_{\theta } (\mathbf{y}, \mathcal{H})$ represents the image recovered by the DiFA Network as shown in Fig.~\ref{fig:framework} (b), and $\mathcal{H} \mathbf{x}^{(1)} $ represents the predicted measurements.

If the measurement process $\mathcal{H}^{T}$ is incomplete, then even in the absence of noise, it is fundamentally impossible to learn a complete MSIs solely from the measurement $\mathbf{y}$, as there is no information in the null space of the measurement process $\mathcal{H}$. Thus, we need to learn more information beyond the range space of their inverse~\cite{52}.

\noindent \textbf{Equivariant Consistency.}
Recently, the EI framework~\cite{49} showed that learning with only measurement data $\mathbf{y}$ is possible with an additional transformation invariant assumption on the signal $\mathcal{X}$. That is, for a certain group of transformations (i.e., $\operatorname{shifts, rotations}$, etc.) $\mathcal{G}=\begin{Bmatrix}g_1,\ldots,g_{|\mathcal{G}|}\end{Bmatrix}$ which are unitary matrices $T_{g}\in \mathcal{G}$, if $\forall \mathbf{x}\in\mathcal{X}$, we have $T_g\mathbf{x}\in\mathcal{X}$ for $\forall g\in\mathcal{G}$ and the sets $T_g\mathcal{X}$ and $\mathcal{X}$ are the same.

With the invariance assumption, the following equations should be met in our method:
\vspace{-2pt}
\begin{align}
    \mathcal{F}_{\theta}(\mathcal{H}T_g\mathbf{x})=T_g\mathcal{F}_{\theta}(\mathcal{H}\mathbf{x}),
\end{align}
for $\forall g\in\mathcal{G}$ and $\forall \mathbf{x}\in\mathcal{X}$. This indicates that the composition $\mathcal{F}_{\theta}\circ \mathcal{H}$ should be transformation invariant.

After obtaining the estimated MSI $\mathbf{x}^{(1)}  = \mathcal{F}_{\theta } (\mathbf{y}, \mathcal{H})$, based on the transformation invariant property, we obtain $\mathbf{x}^{(2)}=T_g\mathbf{x}^{(1)}$ and subsequently feed it to $\mathcal{H}$ and $\mathcal{F}_{\theta}$ as illustrated in Fig.~\ref{fig:framework} (a), resulting in a recovered MSI:
\vspace{-2pt}
\begin{align}
    \mathbf{x}^{(3)}=\mathcal{F}_{\theta}(\mathcal{H}\mathbf{x}^{(2)})=\mathcal{F}_{\theta}(\mathcal{H}T_g\mathbf{x}^{(1)}),
\end{align}
which is the estimation of  $\mathbf{x}^{(2)}$. Thus, our equivariant consistency (EC) loss is formulated as:
\begin{equation}
  \label{EI_LOSS}
    \mathcal{L}_{EC} 
     = ||\mathbf{x}^{(2)} - \mathbf{x}^{(3)}||^{2}\\
    = ||T_{g}\mathcal{F}_{\theta}(\mathbf{y}) - \mathcal{F}_{\theta}(\mathcal{H}(T_{g}\mathcal{F}_{\theta}(\mathbf{y})))||^{2}. 
 \end{equation}

The EC loss in Eq. (\ref{EI_LOSS}) incorporates the transformation invariant prior information of $\mathcal{X}$, allowing us to learn additional information that is beyond the range space of $\mathcal{H}^T$, which is impossible by using the MC loss alone in Eq. (\ref{eq:MC}). 

\noindent\textbf{The Overall Training Loss.} By combining the aforementioned losses, the overall objective function $\mathcal{L}_{Total}$ is:
\vspace{-2pt}
\begin{align}
    \mathcal{L}_{Total}= \mathcal{L}_{MC} + \alpha \mathcal{L}_{EC} + \beta \mathcal{L}_{{DISTILL}},
    \label{eq:total}
\end{align}
where $\alpha$, and $\beta$ are hyper-parameters. The pre-trained network $g_{\theta}$ and the one-step diffusion $f_{\theta}$ are jointly trained by minimizing Eq. (\ref{eq:total}) and the teacher network $F_{\theta}$ is frozen.

\begin{table*}[!t]
  \small
  \centering  \caption{PSNR/SSIM of different methods on three zero-shot datasets. The average results of ten images are reported.}
  \vspace{-10pt}
  \setlength{\tabcolsep}{5.5mm}
  \resizebox{\textwidth}{!}{
    \begin{tabular}{c|c|c|p{5.5em}|p{5.5em}|p{5.5em}}
    \toprule
     \rowcolor{gray!30}Methods & Category & Reference & \multicolumn{1}{c|}{ICVL} & \multicolumn{1}{c|}{NTIRE} & \multicolumn{1}{c}{Harvard} \\
    \midrule
    $\lambda$-Net~\cite{miao2019net} & CNN   & ICCV 2019 & 26.62 / 0.763 & 24.20 / 0.656 & 26.84 / 0.658 \\
    ADMM-Net~\cite{ma2019deep} & Deep Unfolding & ICCV 2019 & 26.44 / 0.775 & 25.47 / 0.748 & 28.22 / 0.741 \\
    TSA-Net~\cite{meng2020end} & CNN   & ECCV 2020 & 24.15 / 0.592 & 21.63 / 0.577 & 19.92 / 0.408 \\
    DIP-HSI~\cite{47} & PnP   & ICCV 2021 & 31.53 / 0.832 & 28.71 / 0.829 & 27.56 / 0.712 \\
    CST-L+~\cite{cai2022coarse} & Transformer & ECCV 2022 & 25.68 / 0.773 & 24.39 / 0.768 & 27.75 / 0.724 \\
    BiSRNet~\cite{cai2023binarized} & BNN   & NeurIPS 2023 & 24.71 / 0.707 & 22.49 / 0.637 & 25.59 / 0.624 \\
    RDLUF-MixS2-9stg~\cite{dong2023residual} & Deep Unfolding & CVPR 2023 & 35.89 / 0.945 & 32.17 / 0.909 & 31.04 / 0.820 \\
    MST-L~\cite{cai2022mask} & Transformer & CVPR 2022 & 27.38 / 0.804 & 25.24 / 0.775 & 27.49 / 0.706 \\
    \rowcolor{lightgray}\textbf{MST-DiFA} & \textbf{Diffusion} & \textbf{Ours} & \textbf{32.81 / 0.894} & \textbf{30.86 / 0.880} & \textbf{30.31 / 0.796} \\
    HDNet~\cite{hu2022HDNet} & Transformer & CVPR 2022 & 25.71 / 0.749 & 24.16 / 0.749 & 26.99 / 0.687 \\
    \rowcolor{lightgray}\textbf{HDNet-DiFA} & \textbf{Diffusion} & \textbf{Ours} & \textbf{30.62 / 0.822} & \textbf{28.08 / 0.784} & \textbf{28.15 / 0.700} \\
    DAUHST-9stg~\cite{cai2022degradation} & Deep Unfolding & NuerIPS 2022 & 30.21 / 0.886 & 29.08 / 0.878 & 30.61 / 0.809 \\
    \rowcolor{lightgray}\textbf{DAUHST-DiFA} & \textbf{Diffusion} & \textbf{Ours} & \textbf{37.19 / 0.952} & \textbf{34.71 / 0.939} & \textbf{31.93 / 0.837} \\
    PADUT-3stg~\cite{li2023pixel} & Deep Unfolding & ICCV 2023 & 29.43 / 0.866 & 27.74 / 0.839 & 28.94 / 0.768 \\
    \rowcolor{lightgray}\textbf{PADUT-DiFA} & \textbf{Diffusion} & \textbf{Ours} & \textbf{35.07 / 0.919} & \textbf{32.51 / 0.897} & \textbf{30.33 / 0.793} \\
    DPU-9stg~\cite{zhang2024dual} & Deep Unfolding & CVPR 2024 & 33.76 / 0.934 & 32.28 / 0.920 & 31.64 / 0.829 \\
    \rowcolor{lightgray}\textbf{DPU-DiFA} & \textbf{Diffusion} & \textbf{Ours} & \textbf{36.39 / 0.918} & \textbf{33.25 / 0.927} & \textbf{32.15 / 0.849} \\
    SSR-L~\cite{zhang2024improving} & Transformer & CVPR 2024 & 36.27 / 0.947 & 34.10 / 0.938 & 31.34 / 0.823 \\
    \rowcolor{lightgray}\textbf{SSR-DiFA} & \textbf{Diffusion} & \textbf{Ours} & \textbf{37.42 / 0.941} & \textbf{37.14 / 0.964} & \textbf{32.46 / 0.857} \\
    \bottomrule
    \end{tabular}%
  \label{tab:average}%
}
\end{table*}%

\begin{table}[t]
  \centering
  \small
  \setlength{\tabcolsep}{4mm}
  \caption{Inference time (a 256 $\times$ 256 MSI) and average reconstruction quality (PSNR/SSIM over 10 images) of diffusion-based methods on zero-shot datasets.}
  \vspace{-8pt}
  \resizebox{0.45\textwidth}{!}{
  \begin{tabular}{c|c|c|c|c|c}
  \toprule
  \rowcolor{gray!30}
  \multicolumn{2}{c|}{Methods} & DiffSCI & LADE-10stg & PSR-SCI-D & SSR-DiFA \\
  \midrule
  \multicolumn{2}{c|}{Reference} & CVPR 2024 & ECCV 2024 & ICLR 2025 & \textbf{Ours} \\
  \midrule
  \multirow{2}{*}{\shortstack{Inference\\Complexity}} 
  & Step  & 50    & \underline{8} & 50    & \textbf{1} \\
  & Time  & 84.54s & \underline{1.10s} & 8.90s & \textbf{0.22s} \\
  \midrule
  \multirow{3}{*}{\shortstack{Quality\\Index}} 
  & ICVL  & 33.02/0.868 & 35.89/0.904 & \underline{37.14/0.918} & \textbf{37.42/0.941} \\
  
  & NTIRE & 32.76/0.903 & 33.58/0.923 & \underline{35.53/0.948} & \textbf{37.14/0.964} \\
  & Harvard & 24.68/0.628 & 28.02/\underline{0.739} & \underline{29.02}/0.728 & \textbf{32.46/0.857} \\
  \bottomrule
  \end{tabular}%
  }
  \label{tab:inference}%
\end{table}

\begin{figure*}[t]
    \centering    \includegraphics[width=\linewidth]{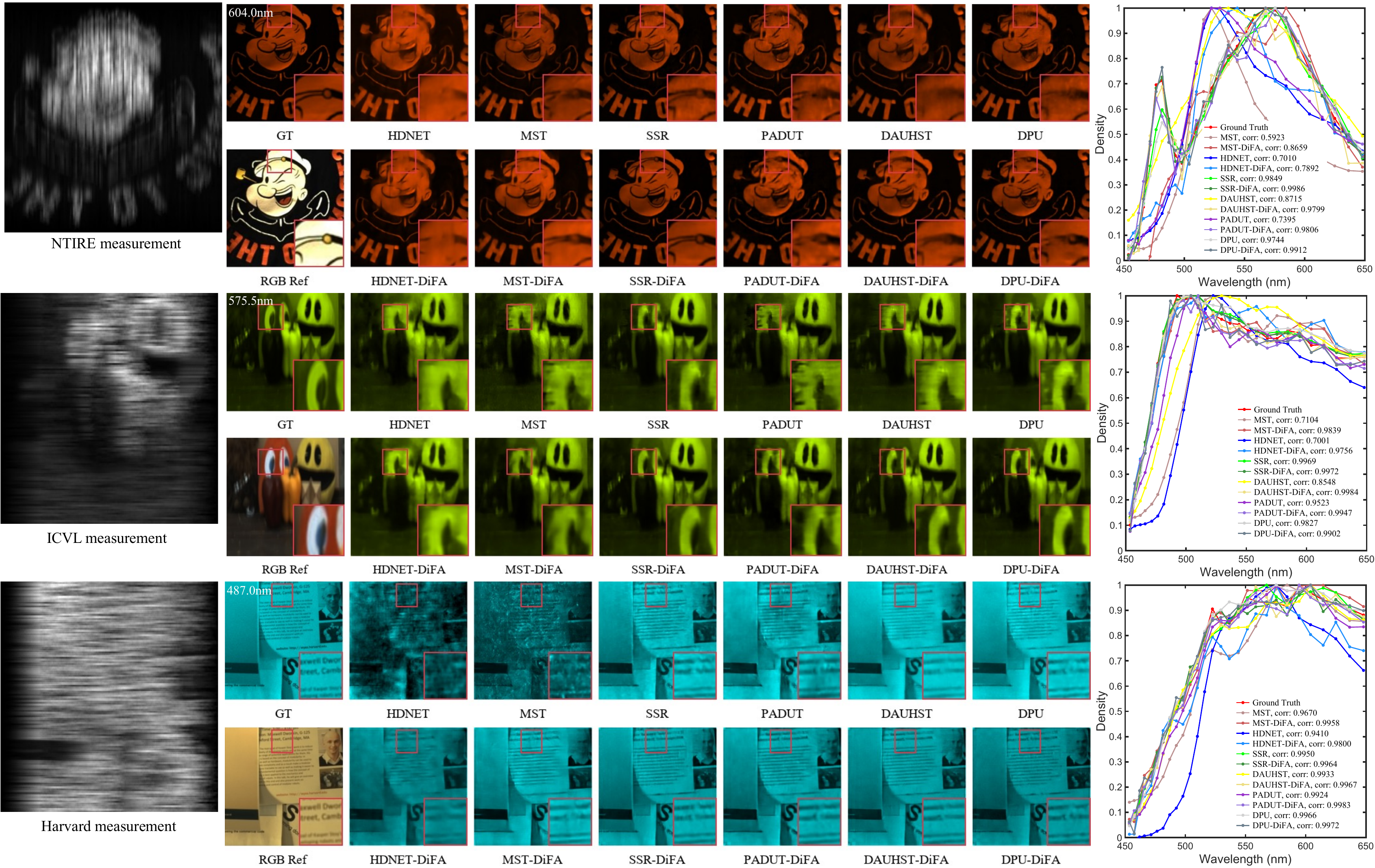}
    \vspace{-7mm}
    \caption{Simulated MSI reconstruction comparisons on zero-shot datasets NTIRE, ICVL and Harvard. The right shows the reconstructed spectral curves corresponding to the selected region.}
    \label{fig:vis}
\end{figure*}

\section{Experiments}
\subsection{Experiment Setting}
\label{sec:exp}
\noindent\textbf{Simulated Dataset.} Three benchmark MSI datasets NTIRE~\cite{arad2022ntire},ICVL~\cite{arad2016sparse}, and Harvard~\cite{choi2017high} are used here. Following~\cite{meng2020end}, we obtain the simulated 2-D compressed measurement $\mathbf{y}$ for the three datasets. In the experiments, the number of bands is 28 and the image size is $256\times 256$ for NTIRE, ICVL, and Harvard. The exact data splits are provided in the code repo under data/*.list.

\noindent\textbf{Real Dataset.} 
Five real MSIs collected by the CASSI system~\cite{meng2020end} are used in the experiments.

\noindent\textbf{Implementation Detail.} The number of training steps is set to 50,000. And we set $\alpha$ = 1, and $\beta$ = 0.001 in Eq. (\ref{eq:total}). The denoising network \( f_{\theta} \) for one-step diffusion is a U-Net incorporating several Swin Transformer blocks~\cite{yue2024efficient}. 
An Nvidia RTX 4090 GPU is used for model training. Our method is implemented by PyTorch.

\begin{figure}[t]
    \centering
    \includegraphics[ width=\linewidth]{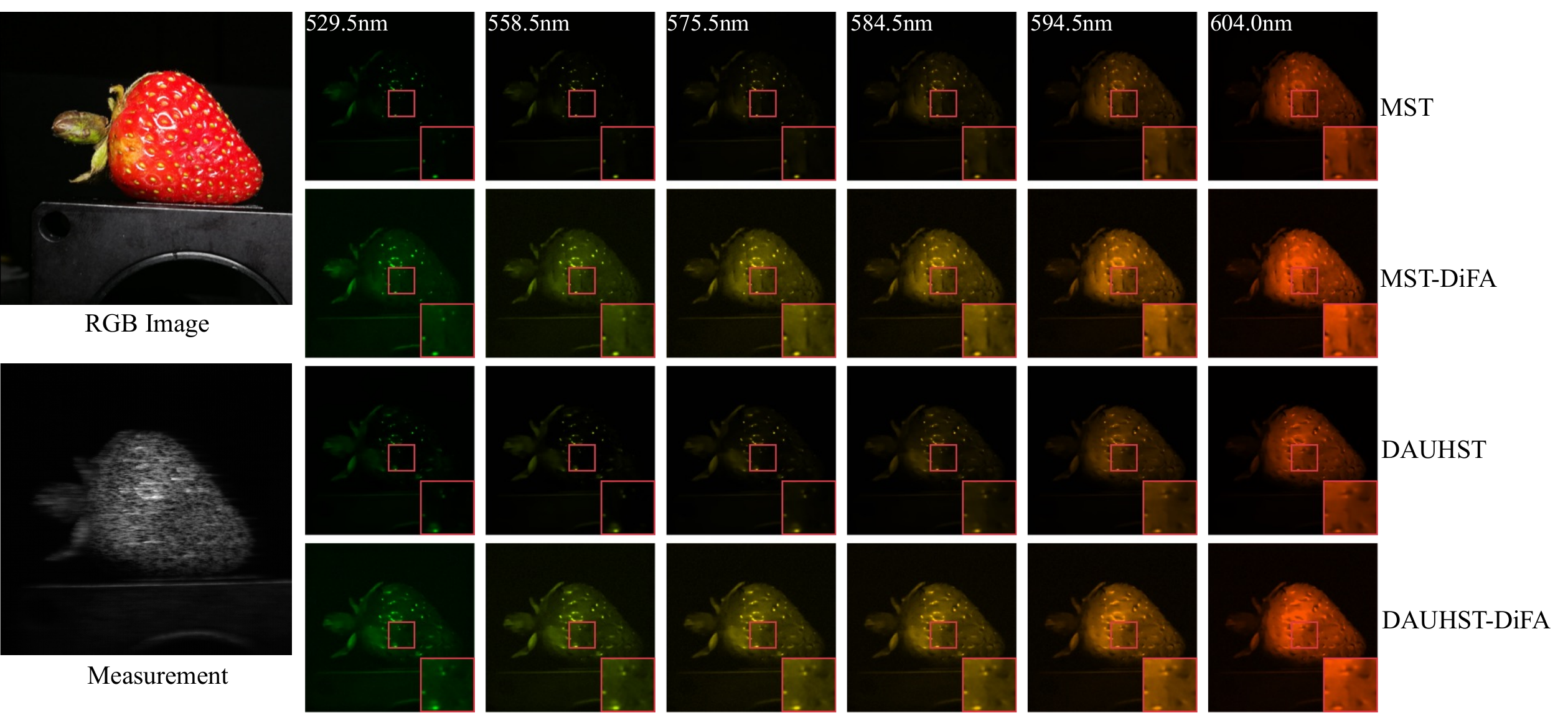}
    \vspace{-5mm}
    \caption{Real MSI reconstruction results on Scene 4 with 6 spectral channels. Please zoom in for a better view.}
    \label{fig:real}
    \vspace{-5pt}
\end{figure}

\noindent\textbf{Compared Methods.} 
We compare our method with 15 state-of-the-art methods, including the end-to-end networks ($\lambda$-Net~\cite{miao2019net}, TSA-Net~\cite{meng2020end}, BiSRNet~\cite{cai2023binarized}, CST-L~\cite{cai2022coarse}, MST-L~\cite{cai2022mask}, HDNet~\cite{hu2022HDNet} and SSR-L~\cite{zhang2024improving}), the PnP networks (DIP-MSI~\cite{47}), the deep unfolding methods (ADMM-Net~\cite{ma2019deep}, DAUHST-9stg~\cite{cai2022degradation}, PADUT-3stg~\cite{li2023pixel}, RDLUF-MixS2-9stg~\cite{dong2023residual} and DPU-9stg~\cite{zhang2024dual}) and diffusion-based methods (DiffSCI~\cite{pan2024diffsci}, LADE-10stg~\cite{wu2025latent} and PSR-SCI-D~\cite{zeng2025spectral}). Since our method is flexible and can incorporate existing end-to-end and deep unfolding SCI reconstruction neural networks as the initial predictor, we employ the MST-L, HDNet, SSR-L, DAUHST-9stg, PADUT-3stg, and DPU-9stg networks that are trained on CAVE dataset as our initial predictor and refer to our methods as MST-DiFA, HDNet-DiFA, SSR-DiFA, DAUHST-DiFA, PADUT-DiFA and DPU-DiFA, respectively. Note that these pre-trained networks were not trained on the test datasets (ICVL, NTIRE, and Harvard), and all experiments are conducted in the zero-shot setting.

\noindent\textbf{Evaluation Metrics.} We evaluate the performance of different methods with the peak signal-to-noise ratio (PSNR) and the structural similarity index metrics (SSIM).
%~\cite{wang2004image}. 
%-------------------------------------------------------------------------
\vspace{-2pt}
\subsection{Quantitative Results}
We show the quantitative results of different methods on ICVL, NTIRE and Harvard datasets in Tab. \ref{tab:average} and \ref{tab:inference}. Our method SSR-DiFA achieves the best performance in terms of PSNR and SSIM on all three datasets. It is observed in Tab. \ref{tab:average} that all our methods outperform their counterparts. Specifically, the PSNR improvements on ICVL are 5.43dB for MST-DiFA over MST-L,  6.98dB for DAUHST-DiFA over DAUHST-9stg, 5.64dB for PADUT-DiFA over PADUT-3stg. This indicates that our method is highly generalized to varying initial predictor.

In Tab.~\ref{tab:inference}, it is observed that compared with the latest SOTA methods DiffSCI, LADE-10stg and PSR-SCI-D, which are the current leading diffusion-based methods in SCI, we obtain the best performance on all three datasets, demonstrating the effectiveness of our design. Compared with LADE-10stg and PSR-SCI-D, which retrain or fine-tune pre-trained diffusion models with paired MSIs, we train a one-step diffusion model from scratch in a self-supervised manner, which only uses 2-D measurement for training, which are more applicable in real applications. Moreover, our method yields the fastest inference speed of 0.22 seconds among the diffusion-based methods, demonstrating the efficiency of our approach.

\begin{table}[t]
  \centering
  \large
  \caption{Ablation study on different components (top) and OSD (bottom) on zero-shot dataset NTIRE.}
  \vspace{-11pt}
  \begin{subtable}[h]{0.45\textwidth}
    \setlength{\tabcolsep}{2mm}
    \centering
    \resizebox{\textwidth}{!}{
        \begin{tabular}{c|c|ccc|cccc}
        \toprule
        \rowcolor{gray!30}
        Case  & DiFA  & MC    & EC    & DSTILL & HDNet-DiFA & MST-DiFA  & PADUT-DiFA & DAUHST-DiFA \\
        \midrule
        \textbf{(a)} & ×     & ×     & ×     & ×     & 24.16 / 0.749 & 25.24 / 0.775  & 27.74 / 0.839 & 29.08 / 0.878 \\
        \textbf{(b)} & \checkmark     & \checkmark     & ×     & ×     & 26.41 / 0.765 & 26.97 / 0.800 & 28.23 / 0.848 & 29.75 / 0.889 \\
        \textbf{(c)} & \checkmark     & \checkmark     & ×     & \checkmark     & 26.82 / 0.779 & 27.05 / 0.801 & 28.60 / 0.851 & 29.81 / 0.890 \\
        \textbf{(d)} & \checkmark     & \checkmark     & \checkmark     & ×     & \underline{28.15 / 0.782} & \underline{30.52 / 0.868} & \underline{32.44 / 0.909} & \underline{34.50 / 0.937} \\
        \textbf{(e)} & \checkmark     & \checkmark     & \checkmark     & \checkmark     & \textbf{28.08} / \textbf{0.784} & \textbf{30.86 / 0.880} & \textbf{32.51 / 0.897} & \textbf{34.71 / 0.939} \\
        \bottomrule
        \end{tabular}
      }
    \label{subtable:2}
  \end{subtable}
  \begin{subtable}[h]{0.45\textwidth}
    \setlength{\tabcolsep}{5mm}
    \centering
      \resizebox{\textwidth}{!}{
        \begin{tabular}{c|cccc}
        \rowcolor{gray!30}
        \toprule
        Methods & HDNet-DiFA & MST-DiFA  & PADUT-DiFA & DAUHST-DiFA \\
        \midrule     
        Baseline & 24.16 / 0.749 & 25.24 / 0.775 & 27.74 / 0.839 & 29.08 / 0.878 \\
        w/o OSD & \underline{27.42} / \textbf{0.810} & \underline{28.83 / 0.834} & \underline{31.96 / 0.903} & \underline{33.11 / 0.914} \\
        with OSD & \textbf{28.08} / \underline{0.784} & \textbf{30.86 / 0.880} & \textbf{32.51 / 0.897} & \textbf{34.71 / 0.939} \\
        \bottomrule
        \end{tabular}
      }
    \label{subtable:2}
  \end{subtable}
  \label{table:main}
\end{table}

\vspace{-6pt}
\subsection{Qualitative Results}

We show the visual results of different methods on the simulated datasets NTIRE, ICVL, and Harvard in Fig.~\ref{fig:vis}. We can see that our DiFA provides better visual results with more details and fewer artifacts than the original methods. Moreover, we show the spectral curves of different methods on the right of Fig.~\ref{fig:vis}. It is observed that our DiFA strategy effectively improves the spectral reconstruction accuracy, demonstrating the superior spectral reconstruction ability of DiFA. In addition, we show the visual results of MST, DAUHST and our methods MST-DiFA and DAUHST-DiFA on the real data set in Fig.~\ref{fig:real}. Compared with the original methods, our approaches obtain improved visual quality as demonstrated in the region of strawberry.

\vspace{-2pt}
\subsection{Experimental analysis}
\subsubsection{4.4.1 Ablation Study.}
There are three key components in our method, including DiFA, the EC loss $\mathcal{L}_{EC}$ and the distillation loss $\mathcal{L}_{DISTILL}$. In addition, we investigate the influence of the data fidelity MSE loss $\mathcal{L}_{MC}$. We remove different components from our methods and conduct experiments on NTIRE with four different initial predictors HDNet, MST, PADUT and DAUHST. The results are shown in Tab.~\ref{table:main} top. Case (a) represents the original method. It is observed that compared with the baselines, i.e., case (a), all other cases obtain improved performance. Our final methods in case (e) with all components yield the best performance, demonstrating the effectiveness of each design. 
\vspace{-2pt}
\subsubsection{4.4.2 The Influence of the One-step Diffusion in DiFA.}
The OSD model in DiFA is used to estimate the details of MSI for the refinement of the initial predicted MSI. To verify its effectiveness, we conduct experiments on zero-shot dataset NTIRE by removing the one-step diffusion from our approach. Our four methods with different initial predictors are tested. The results are shown in Tab.~\ref{table:main} bottom. It is observed that without using OSD, the PSNR performance is dropped. For instance, the PSNR is reduced by 2.03dB for MST-DiFA. It should be noted that compared with the baselines HDNet, MST-L, PADUT-3stg and DAUHST-9stg, our reduced methods without OSD still obtain significant performance improvement.

\section{Conclusion}
In this paper, we propose a novel self-supervised one-step diffusion refinement framework for SCI, which reconstructs a high-quality MSI with an initial predictor and an OSD model. The OSD model provides details for the rough prediction obtained by the initial predictor, leading to a high-quality MSI with fine details. We design a distillation loss in the RGB space for knowledge transfer from the RGB domain to the MSI domain and devise a self-supervised learning method based on a transformation invariant property, which allows effective model training with only 2-D measurement. Our method achieves state-of-the-art performance on both simulated and real data, offering faster inference and excellent generalization compared to the latest diffusion-based methods in the field.

\section{Acknowledgments}
This work was supported in part by the National Natural Science Foundation of China under grants 42301425 and T2525018, in part by the China Postdoctoral Science Foundation (2023M743299), in part by the ``CUG Scholar'' Scientific Research Funds at China University of Geosciences (Wuhan) under grant 2022164, and in part by the Fundamental Research Funds for the Central Universities, China University of Geosciences (Wuhan) under grant CUG240628.

\bibliography{aaai2026}
\end{document}